\documentclass[ referee, usenatbib]{mnras}
\usepackage{epsfig}
\usepackage{newtxtext,newtxmath}
\usepackage{bm}

\usepackage[T1]{fontenc}
\usepackage{ae,aecompl}

\usepackage{graphicx}	
\usepackage{amsmath}	
\usepackage{amssymb}	

\title[Evolution of Nearly Polar Low Mass Circumbinary Discs]{Linear Analysis of the Evolution of Nearly Polar Low Mass Circumbinary Discs}

\author[Lubow \& Martin]{Stephen
  H. Lubow$^1$ \thanks{E-mail: lubow@stsci.edu} and Rebecca G. Martin$^{2}$  \thanks{E-mail: rebecca.martin@unlv.edu}\\ 
$^1$Space Telescope Science Institute, 3700 San Martin
  Drive, Baltimore, MD 21218, USA \\
  $^{2}$Department of Physics and Astronomy, University
  of Nevada, Las Vegas, 4505 South Maryland Parkway, Las Vegas, NV
  89154, USA \\ }

\date{Accepted XXX. Received YYY; in original form ZZZ}

\pubyear{2017}

\begin{document}

\label{firstpage}
\pagerange{\pageref{firstpage}--\pageref{lastpage}}
\maketitle


\begin{abstract}
\cite{Martin2017} showed through simulations that an initially tilted  disc around an eccentric
binary can evolve to polar
alignment in which the disc lies perpendicular to the binary orbital plane. We apply linear theory
to show both analytically and numerically that a nearly  polar aligned low mass circumbinary disc evolves to polar alignment
and determine the alignment timescale.  Significant disc evolution towards the polar state around moderately
eccentric binaries can occur for typical protostellar disc parameters in less than a typical disc
lifetime for binaries with orbital periods of order 100 years or less.
Resonant torques are much less effective at truncating the inner parts of circumbinary polar discs than the inner parts of coplanar discs.
For polar discs, they vanish for a binary eccentricity of unity.
The results agree with the simulations in showing that discs can evolve to a polar state. Circumbinary planets may then form in such discs and reside on polar orbits.
\end{abstract}




\section{Introduction}
\label{intro}

Observations of protostellar circumbinary discs have revealed misalignments between
the orbital planes of the binaries and their discs
in some cases. For example, the
peculiar light curve of KH 15D can be explained by a misaligned
precessing disc \citep[e.g.][]{Winn2004, Chiang2004,Capelo2012}. The
binary protostar IRS~43 has a misalignment greater than $60^\circ$
between the binary and the disc \citep{Brinch2016}.
Binary GG Tau may be misaligned by $25^\circ-30^\circ$ from its
circumbinary disc 
\citep{Cazzoletti2017}.
The debris disc
around 99 Herculis is most likely perpendicular to the binary
orbital plane \citep{Kennedy2012}.  

There are some mechanisms that could explain how misalignments could arise between circumbinary discs
and their central binary orbital planes.  
Turbulence in a star-forming gas cloud
could result
in disc misalignments with the central binary  \citep{Offner2010,Tokuda2014, Bate2012}. 
  If a young binary star system accretes
material after its formation process, the material is likely to be
misaligned with the binary orbit and so misaligned discs may form
around young binary stars \citep{Bateetal2010}.
In addition, misalignment can be the
result of binary star formation from an elongated cloud whose axes are
misaligned with respect to the cloud rotation axis
\citep[e.g.][]{Bonnell1992}.  Such misalignment mechanisms likely
apply to both circumstellar and circumbinary discs.

The Kepler telescope has detected about a dozen
extra-solar circumbinary planets \citep[e.g.,][]{Doyle2011, Welsh2015,Kostov2016}. Such planets likely
formed in a circumbinary disc. The planets
found to date are nearly coplanar with the binary
orbit. It would be useful to know if there are
possible long lived misaligned circumbinary discs that
could result in the formation of noncoplanar circumbinary planets.

Previous analytic and numerical studies of the tilt evolution of circumstellar and circumbinary discs 
have largely focused on the alignment process  towards the binary orbital plane
\citep[e.g.,][]{PT1995, LP1997, Lubow2000, Nixonetal2011b, Facchinietal2013, Lodato2013, Foucart2013, Foucart2014}.
It is natural to expect that tilted discs will evolve towards alignment with the binary
since such a configuration would have minimum energy for a fixed density distribution.
In presence of dissipation, one would expect evolution to coplanar alignment with the binary orbit.  A notable
exception is the study by \cite{Aly2015} who found that disc alignment can occur perpendicular to the binary orbital plane in the context
of a cool disc that undergoes tearing around  a supermassive black hole binary.
We are interested in warmer discs that are found in the protostellar/protoplanetary context.

Binary-disc resonances can play a role in causing a circumstellar disc to not achieve a coplanar configuration
\citep{Lubow1992, PT1995, Lubow2000}. However, these effects are either weak or are restricted
to certain disc parameters.

Recently, we showed through smoothed particle hydrodynamic (SPH) simulations that a  somewhat misaligned low mass
protostellar circumbinary disc can evolve to an apparently stable polar state in which the disc plane is perpendicular
to the orbital plane of the binary \citep{Martin2017}. A key ingredient is that the binary
be on an eccentric orbit. In this polar orientation, the rotation axis of the disc is
parallel to the eccentricity vector of the binary.


A misaligned test (massless) particle orbit around a circular orbit binary undergoes
nodal precession. The angular momentum vector of the test particle
precesses about the binary angular momentum vector. Similarly, a misaligned
protostellar disc around a circular orbit binary can undergo the same form of precession. A protostellar  disc
is often able to hold itself together and precess as a solid body 
\citep[e.g.,][]{PT1995, LP1997,Lubow2000, Foucart2014}. Differential precession across a disc can cause it to become
highly warped. But
protostellar discs are often warm enough for
global radial communication to occur through pressure induced bending waves that permit
the disc to resist warping. Dissipation within the disc leads to
alignment with the binary orbital plane.

For eccentric orbit binaries, test  particle orbits behave somewhat differently.
Instead of precessing about the orbital axis of the binary, they can 
instead precess about the eccentricity vector of the binary  that lies perpendicular
to the binary angular momentum axis \citep{Farago2010,Doolin2011,LZZ2014}.
There is a critical initial inclination above which the  transition occurs
from precessing about the orbital axis of the binary to precessing about the eccentricity vector of the binary. 
The critical misalignment angle
depends upon the eccentricity of the binary and the longitude of the ascending node of the particle's orbit.
Polar alignment of a moderately misaligned low mass disc can be understood in terms of 
transitions between these precessing orbits to a nonprecessing polar state
as the disc undergoes tidal dissipation \citep{Martin2017}.

The stability of nearly polar discs or rings 
has been explored in the context of galaxies and planets.
Polar rings are observed around galaxies  \citep{Schweizer1983, Combes2014, Iodice2015}. 
Their frequency around galaxies is estimated to be $\sim 5\%$, after various selection effects are taken into account \citep{Whitmore1990}. There are  several models for the  origin of
polar rings in galaxies that generally involve the gravitational effects of dark matter halos acting
on infalling material (gas or galaxies).  In one early model, the ring attains a polar
configuration  through the effects of a spheroidal  or triaxial potential acting
on a dissipative gaseous ring \citep{Durisen1983, Steiman1990}.  Such a configuration 
is similar to the eccentric binary case, in which the time-averaged potential
has a triaxial form \citep{Farago2010}. In a completely different context,
\cite{Dobrovolskis1989} showed that
a possible polar ring around Neptune would be stable through similar arguments.

In this paper, we  analyze the dynamics of a nearly  polar  circumbinary disc about an eccentric binary.
A tilted protostellar disc is subject to  the development of bending waves
that are communicated by pressure which has not been previously considered in models of polar discs.  In addition, Keplerian discs experience a near resonance
of horizontal epicyclic motions  driven by horizontal  pressure gradients  in a warped disc
\citep{Papaloizou1983}. 

As we were completing the work on this paper, a preprint by \cite{Zanazzi2017} appeared on this topic.
Our paper has a somewhat different emphasis. Both papers agree that the polar state is stable and obtain
similar tilt evolution timescales.

The outline of the paper is as follows. Section 1 contains the introduction.
Section 2 describes the linear equations used
to model the evolution of a nearly polar disc.
Section 3 describes solutions to those equations
based on an expansion in which the tidal potential is weak.
Section 4 discusses numerically determined modes for
nearly polar discs.
Section 5 analyzes the tidal truncation of the inner parts of a polar circumbinary disc.
Section 6 briefly discusses the stability of a disc that
that is perpendicular to the binary orbital plane and the polar disc plane.
Section 7 discusses some energetics issues related to
polar discs. Section 8 contains a summary.

\section{Nearly Polar Disc Model}
\label{sec:model}

We consider an eccentric binary with component stars that have masses $M_1$ and $M_2$ and
total mass $M=M_1+M_2$ in an orbit with semi--major axis $a_{\rm b}$ and eccentricity $e_{\rm b}$.  
We are interested in nearly polar, low mass circumbinary discs. Such discs have their angular momentum vectors nearly parallel
(or antiparallel) to their binary eccentricity vectors that point from the binary centers of mass towards their binary pericenters.
In this configuration, the disc is nearly perpendicular the binary orbital plane.
To describe this configuration, we define a Cartesian coordinate system $(x,y,z)$ whose origin is at the binary center of mass and with the $y$-axis parallel to the binary angular momentum $\bm{J}_{\rm b}$
and the $z$-axis parallel to the binary eccentricity vector $\bm{e}_{\rm b}$. 
We consider the disc to be 
composed of circular rings with surface density $\Sigma(r)$ whose  orientations
vary with radius $r$ and orbit with Keplerian angular speed $\Omega(r)$. We denote the unit vector parallel to the ring angular  momentum at each radius by $(\ell_x,\ell_y, \ell_z)$. We consider small departures of the disc from the $x-y$ plane,
so that
$| \ell_x| \ll 1, |\ell_y| \ll 1,$ and $\ell_z  \approx 1$.

We apply equations~(12) and (13)
in \cite{Lubow2000} for the evolution of the disc 2D tilt vector $\bm{\ell}(r,t)=(\ell_x, \ell_y)$ and 2D internal torque $\bm{G}(r,t)=(G_x,G_y)$.  The tilt evolution equation is given by
\begin{equation}
\Sigma r^2 \Omega \frac{\partial \bm{\ell}}{\partial t}=\frac{1}{r}\frac{\partial \bm{G}}{\partial r}+\bm{T},
\label{lubow1}
\end{equation}
where $\bm{T}$ is the tidal torque per unit area due to the eccentric binary whose orbit lies in the $x-z$ plane.
Equation~(13) in \cite{Lubow2000} provides the internal torque evolution equation
\begin{equation}
\frac{\partial \bm{G}}{\partial t} - \omega_{\rm a} \bm{e}_z \times \bm{G}+\alpha \Omega  \bm{G}=\frac{{\cal I} r^3 \Omega^3 }{4}\frac{\partial \bm{\ell}}{\partial r},
\label{lubow2}
\end{equation}
where $\alpha$ is the usual turbulent viscosity parameter, $\omega_{\rm a}(r)$ is the apsidal precession rate for a disc that is polar that is given in Equation~(\ref{opp}) of Appendix A, and
\begin{equation}
{\cal I} = \int \rho z^2 dz,
\end{equation}
for disc density $\rho(r)$.
We apply boundary conditions that the internal torque vanishes at the inner and outer disc edges $r_{\rm in}$ and $r_{\rm out}$,
respectively. That is,
\begin{equation}
\bm{G}(r_{\rm in}, t) = \bm{G}(r_{\rm out}, t) =0.
\label{GBC}
\end{equation} 
 This is a natural boundary condition because the internal torque vanishes just outside the disc boundaries. Thus, any smoothly
varying internal torque would need to satisfy this condition.

The torque term due to the eccentric binary  follows from an application of
  equations (2.17) and (2.18) in \cite{Farago2010}. Note that we apply
a different coordinate  system in which
our $(x,y,z)$ coordinates corresponds to their  $(y, z, x)$ coordinates. The torque term is
expressed as 
\begin{equation}
\bm{T}=\Sigma r^2 \Omega \bm{\tau}
\label{Ttor}
\end{equation}
with 
\begin{equation}
\bm{\tau}=(a(r) \ell_y, b(r) \ell_x) 
\label{tau}
\end{equation}
and
\begin{equation}
a(r) =-  (1+4e_{\rm b}^2) \, \omega_{\rm p}(r), 
\label{taux}
\end{equation}
and
\begin{equation}
b(r)=5  e_{\rm b}^2 \, \omega_{\rm p}(r),
\label{tauy}
\end{equation}
where frequency $\omega_{\rm p}$ is given by
\begin{equation}
\omega_{\rm p}(r)=\frac{3}{4}\frac{\beta}{M} \left(\frac{a_{\rm b}}{r} \right)^{7/2} \Omega_{\rm b},
\label{omegap}
\end{equation}
and $\beta$ is the reduced mass 
\begin{equation}
\beta=\frac{M_1M_2}{M_1+M_2}.
\end{equation}

We seek solutions of the form $\bm{\ell} \propto e^{i\omega t}$ and $\bm{G} \propto e^{i\omega t}$ and Equations (\ref{lubow1}) and (\ref{lubow2}) become
\begin{equation}
 i  \omega   \Sigma r^2 \Omega \bm{\ell} =\frac{1}{r}\frac{d \bm{G}}{d r}+  \Sigma r^2 \Omega\bm{\tau}
\label{l1}
\end{equation}
and
\begin{equation}
i \omega \bm{G} - \omega_{\rm a} \bm{e}_z \times \bm{G}+\alpha \Omega  \bm{G}=\frac{{\cal I} r^3 \Omega^3}{4} \frac{d \bm{\ell}}{d r},
\label{l2}
\end{equation}
respectively.

\section{Nearly Rigid Disc Expansion}
\label{sec:expansion}
\subsection{Lowest order}

We apply the nearly rigid tilted disc expansion procedure in \cite{Lubow2000}.
We  expand variables in the tidal potential that is considered to be weak as follows:
\begin{eqnarray}
a &=& A^{(1)},\\
b &=& B^{(1)}, \nonumber \\
\bm{\ell} &=& \bm{\ell}^{(0)} + \bm{\ell}^{(1)} + \cdots,  \nonumber \\
\omega &=& \omega^{(1)} + \omega^{(2)} + \cdots,  \nonumber \\
\bm{G} &=& \bm{G}^{(1)} + \bm{G}^{(2)} + \cdots,  \nonumber \\
\bm{\tau} &=& \bm{\tau}^{(1)} + \bm{\tau}^{(2)} + \cdots,   \nonumber
\end{eqnarray}
where $a$ and $b$ are given by Equations (\ref{taux}) and  (\ref{tauy}), respectively.  $a$ and $b$ depend on the tidal
potential and are regarded as first order quantities.

To lowest order, the disc is rigid and the tilt vector $\bm{\ell}^{(0)} = (\ell_{x}^{(0)}, \ell_{y}^{(0)})$ is constant in radius.
We integrate $r$ times Equation (\ref{l1}) over the entire disc and apply the boundary conditions given by Equation (\ref{GBC}) to obtain
\begin{equation}
\int_{r_{\rm in}}^{r_{\rm out}} \Sigma r^3 \Omega(i\omega^{(1)} \bm{\ell}^{(0)}- \bm{\tau}^{(1)})\, dr=0,
\end{equation}
where  
\begin{equation}
\bm{\tau}^{(1)}(r) = (A^{(1)}(r) \, \ell_{y}^{(0)}, B^{(1)}(r)\, \ell_{x}^{(0)}).
\end{equation}
We then obtain for the disc frequency in lowest order 
\begin{equation}
\omega^{(1)} = \frac{3 \sqrt{5}}{4} e_{\rm b} \sqrt{1+4 e_{\rm b}^2} \frac{M_1 M_2}{M^2}\left<\left(\frac{a_{\rm b}}{r} \right)^{7/2} \right>  \Omega_b,
\label{om1}
\end{equation}
where the bracketed term involves the angular momentum weighted average
\begin{equation}
 \left<\left(\frac{a_{\rm b}}{r} \right)^{7/2} \right>= \frac{ \int_{r_{\rm in}}^{r_{\rm out}}\Sigma r^3 \Omega  (a_{\rm b}/r)^{7/2} dr}{ \int_{r_{\rm in}}^{r_{\rm out}} \Sigma r^3 \Omega dr}.
 \label{javg}
\end{equation}
The tilt components are related by
\begin{equation}
\ell^{(0)}_y= - \frac{\sqrt{5}\, i\, e_{\rm b}}{\sqrt{1+4e_{\rm b}^2}} \ell^{(0)}_x.
\label{ly0}
\end{equation}

\begin{figure}
\centering
\includegraphics[width=8.0cm]{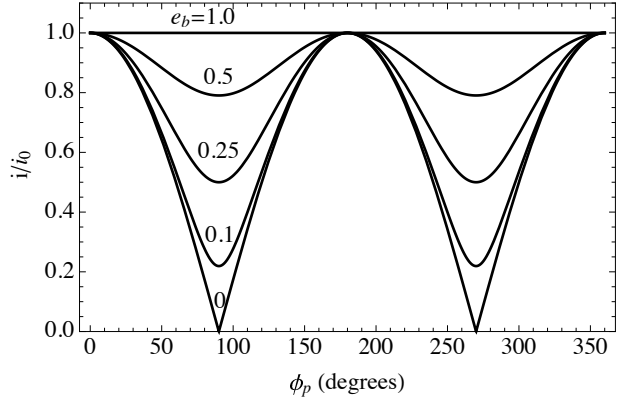}
\caption{Normalized disc tilt angle relative to the polar orientation (Equation (\ref{it})) as a function of precession angle $\phi_{\rm p} = \omega^{(1)} t$ for 
various values of binary eccentricity.  }
\label{ivrst}
\end{figure}

Because $|\ell^{(0)}_x|$ and $|\ell^{(0)}_y|$ differ, the disc undergoes secular tilt oscillations with tilt
variations $i(t)$ with respect to the $x-y$ plane. If we assume the disc is initially tilted along the $x-$axis (i.e., $\ell^{(0)}_x$ is real), then
\begin{equation}
i(t) = i(0) \sqrt{ \frac{1 + 9 e_{\rm b}^2 + (1-e_{\rm b}^2) \cos { ( 2 \omega^{(1)} t )} }{2(1+4e_{\rm b}^2)}}.
\label{it}
\end{equation}
Figure~\ref{ivrst} plots the tilt as a function of precession angle $\phi_{\rm p} = \omega^{(1)} t$ for 
various values of binary eccentricity.
Tilt oscillations occur because the binary potential is nonaxisymmetric around the direction of the binary eccentricity vector
(the $z-$axis).
For $e_{\rm b}=1$, the oscillations are suppressed because the binary orbit is along a line and the potential
is then axisymmetric about the line.

We integrate $r$ times Equation (\ref{l1}) and apply boundary condition (\ref{GBC}) to find the internal torque to lowest order
\begin{equation}
\bm{G}^{(1)}(r)=\int^r_{r_{\rm in}} \Sigma r^3 \Omega(i\omega^{(1)} \bm{\ell}^{(0)}- \bm{\tau}^{(1)})\, dr'.
\label{G0}
\end{equation}
Equation (\ref{l2}) in lowest order is given by 
\begin{equation}
\alpha \Omega  \bm{G}^{(1)}=\frac{1}{4}{\cal I} r^3 \Omega^3 \frac{d \bm{\ell}^{(1)}}{d r}.
\label{l2a}
\end{equation}
This is used to determine $\bm{\ell}^{(1)}$.


\subsection{Higher order} 
\label{sec:higherorder}

To next order in the nearly rigid tilt approximation, we determine the tilt oscillation amplitude decay rate $Im(\omega^{(2)})$. We show  here analytically that the decay rate  is positive, implying
that the disc evolves to a polar state  where  $\bm{\ell} = 0$.
Circumbinary disc dynamics involving an eccentric orbit binary  is different from  the case involving a circular orbit binary analyzed previously \citep[e.g.][]{  Facchinietal2013, Foucart2014} because of the presence of secular tilt oscillations described by Equation (\ref{it}). 

We determine the deviations from the rigid tilt, i.e., the warp $\bm{\ell}^{(1)}=(\ell^{(1)}_x, \ell^{(1)}_y)$, by integrating Equation (\ref{l2a}) to obtain
\begin{equation}
\bm{\ell}^{(1)}(r)= \bm{\ell}^{(1)}(r_{\rm in})+\int^r_{r_{\rm in}} \frac{ 4 \alpha \bm{G}^{(1)}}{{\cal I} r'^3 \Omega^2} \, dr'.
\label{ell1}
\end{equation}
In this linear analysis we are free to specify the value of one component of $\bm{\ell}(r_{\rm in})$. We 
specify the value of $\ell_x(r_{\rm in}) =   \ell^{(0)}_x(r_{\rm in})$ that is taken to be real quantity and consequently
\begin{equation}
 \ell_x^{(1)}(r_{\rm in})=0.
\label{lxin}
\end{equation}

We take the dot product of Equation (\ref{l1}) with $r \bm{\ell}^*$
and integrate in radius over the disc to obtain  
\begin{eqnarray}
 Im(\omega) \int_{r_{\rm in}}^{r_{\rm out}} \Sigma r^3 \Omega |\bm{\ell}|^2  \, dr &=& -\int_{r_{\rm in}}^{r_{\rm out}} \Sigma r^3 \Omega Re \left(\bm{\tau} \cdot \bm{\ell}^{*} \right) \, dr  \nonumber \\
 &+&  \int_{r_{\rm in}}^{r_{\rm  out}}  Re\left( \bm{G}   \cdot \frac{d\bm{\ell}^*}{dr}  \right) dr,
\label{Imom}
\end{eqnarray}
where the second term on the RHS is obtained by an integration by parts and the application
of boundary conditions  (\ref{GBC}).

The first term on the RHS involves the dot product of the tidal torque with the disc tilt. 
The integrand involves term $Re(\bm{\tau}\cdot \bm{l^*})  = (a+b) Re(\ell_x \ell_y^*)$.
For a  purely precessional torque that involves no secular tilt oscillations, as occurs for a  slightly tilted circumbinary
disc  that lies near the orbital plane of a circular orbit binary, we have that
$a(r)=-b(r)$ in Equation (\ref{tau}). It then immediately follows that the tidal torque term involving $Re(\bm{\tau}\cdot \bm{l^*})$ vanishes
in that case.

The contributions of the tidal torque term in Equation (\ref{Imom}) are less clear in 
the case of a disc undergoing secular tilt oscillations, as
occurs in the nearly polar case, where $a$ and $-b$ differ. 
In fact, the contribution
of this term does not vanish to all orders, as we show in Section \ref{sec:omcont}
for numerically computed modes.
It is easy to see that in zeroth order
$Re(\ell^{(0)}_x \ell^{(0)*}_y)$ vanishes  using  Equation (\ref{ly0}) which implies
that  the tidal torque term vanishes in first order.

The internal torque term that appears as the second term on the RHS of Equation (\ref{Imom})
is nonzero in second order, since to lowest order it involves the product $ \bm{G}^{(1)} \cdot d\bm{\ell}^{(1)*}/dr$.
To properly compare the tidal torque term with the internal torque term on the RHS of Equation (\ref{Imom}),
we must analyze the tidal torque term in second order.

We evaluate Equation (\ref{Imom}) to second order as
\begin{eqnarray}  
 Im(\omega^{(2)})\, J_d |\bm{\ell}^{(0)}|^2 &=& -2 \pi \int_{r_{\rm in}}^{r_{\rm out}} \Sigma r^3 \Omega \, Re \left( \left(\bm{\tau} \cdot \bm{\ell}^{*} \right)^{(2)} \right)\, dr  \nonumber \\ 
 &+& 2 \pi \int_{r_{\rm in}}^{r_{\rm  out}}   Re\left( \bm{G}^{(1)} \cdot \frac{d\bm{\ell}^{(1)*}}{dr}  \right) dr,
\label{lam}
\end{eqnarray}
where 
$J_{\rm d}$ is the angular momentum of the disc 
\begin{equation}
J_{\rm d}=2\pi \int_{r_{\rm in}}^{r_{\rm out}} \Sigma r^3 \Omega \, dr
\end{equation}
and
\begin{equation}
 Re\left( \left(\bm{\tau} \cdot \bm{\ell}^{*} \right)^{(2)} \right) = (A^{(1)} + B^{(1)}) Re\left( \ell^{(1)}_x \ell^{(0)*}_y + \ell^{(0)}_x \ell^{(1)*}_y  \right).
 \label{Tl2}
\end{equation}

In Appendix B1 we show that in first order $Re(\ell_x \ell^{*}_y)$
is constant in radius.  We then evaluate this constant at the disc inner edge and show that it is zero in Appendix B2.
Finally,  in Appendix B3 we  show that the RHS of Equation (\ref{Tl2}) is zero at all radii in the disc.
This occurs because $a(r)/b(r)$ is independent of radius.
We then have that the LHS of Equation (\ref{lam}) balances 
 the second term on the RHS. Applying Equation (\ref{l2a}) we obtain
\begin{equation}
 Im(\omega^{(2)}) = \frac{2 \pi}{J_{\rm d} |\bm{\ell}_0|^2} \int_{r_{\rm in}}^{r_{\rm out}} \frac{4 \alpha|\bm{G}^{(1)}|^2 }{{\cal I} r^3 \Omega^2} \, dr.
 \label{om2}
\end{equation}
This relation agrees with Equation (47) of \cite{Lubow2001} that was derived for a nearly coplanar disc that does not undergo secular tilt oscillations.
The tilt amplitude evolution rate $Im(\omega^{(2)})$ then is positive definite which implies that a disc whose orientation is slightly different from polar will  evolve towards the polar state.
{\it Consequently,  we find analytically that low mass polar discs are secularly linearly stable. }

 Consider a set of discs of a fixed structure. That is, the temperature and density variations are fixed in radius with fixed inner and outer radii, but the temperature and density
normalization can vary. For example, if the temperature is a power law in radius, then the exponent in the power law is taken to be
fixed, but the temperature at some reference radius, such as the disc inner edge, can change across discs.
Equation (\ref{om2}) implies that for fixed disc structure, small departures of tilt from the polar state decay
towards this state on a timescale 
\begin{eqnarray}
t_{\rm evol} &\propto& \frac{(H/r)^2 \Omega_{\rm b}}{\alpha (\omega^{(1)})^2 } 
\label{td1} \\
 &  \propto &  \frac{(H/r)^2 M^2}{\alpha e_{\rm b}^2 (1+4 e_{\rm b}^2) M_1^2 M_2^2 \Omega_{\rm b}},
 \label{td2}
\end{eqnarray}
where $H/r$ is the disc aspect ratio at some reference radius that we can take to be the disc inner radius.
The full dependence (not assuming a fixed disc structure) is more complicated because the decay rate is sensitive to the disc inner radius
that depends in turn on the binary mass ratio, $H/r, \alpha,$ and $e_{\rm b}$, as is discussed later in Section \ref{sec:truncation}.

\section{Numerical Determination of  Modes}
\label{numerical}

\subsection{Method}
In this section, we present results of solving the linearized tilt Equations (\ref{l1}) and (\ref{l2})
numerically. We are again interested in the longest lived mode that is nearly a rigid
tilt mode. We compare the numerical results with those of the expansion method in Section \ref{sec:expansion}.

 We numerically determine complex eigenfunctions $\ell(r)$ and $G$ and eigenvalue $\omega$ in Equations (\ref{l1}) and (\ref{l2}) 
 by a shooting method. 
Equations  (\ref{l1}) and  (\ref{l2}) are first order in radius for the $x$ and $y$ component equations. 
They 
are integrated  outward from the inner radius $r=r_{\rm in}$ to the outer boundary $r=r_{\rm out}$.
To integrate these four equations we need to specify the four variables at the inner radius and the eigenvalue $\omega$.
 Two starting conditions are supplied by the boundary condition of Equation (\ref{GBC}), $G_x(r_{\rm in}) = G_y(r_{\rm in})  =0$.
 We can freely specify one component of $\bm{\ell}$ at the inner boundary and set 
$\ell_x(r_{\rm in}) = 1$. The remaining  quantities  are $\ell_y(r_{\rm in})$ and $\omega$ whose values
are adjusted to satisfy the outer boundary condition of Equation (\ref{GBC}), $G_x(r_{\rm out}) = G_y(r_{\rm out})  =0$.
There are infinitely many modes that satisfy the boundary conditions and have different values of $\ell_y(r_{\rm in})$ and $\omega$. To determine the nearly rigid modes,
we apply initial estimates for the shooting method from analytic nearly rigid tilt modes described in Section \ref{sec:expansion} in a manner described later in Section \ref{sec:tiltevol}.
The calculations were performed using Mathematica software.

\subsection{Disc Models} \label{sec:models}

We determine the linear modes for a set of disc models,
all of which involve an equal mass binary.  
We analyze discs whose parameters are listed in Table 1,
where $s$ and $p$ are defined by $T(r) \propto r^{-s}$ and $\Sigma(r) \propto r^{-p}$, respectively.
We regard Model A as a fiducial model and consider single parameter departures from that model in other models.
Models A, B, D, E, and F involve flared discs  that have a disc aspect ratios $H/r \propto r^{1/4}$.
Such a level of flaring (or more) is expected in so-called `passive' protostellar discs, where the heating is dominated
by the contributions from the central stars at all radii, or in active discs on scales greater than $\sim 1$ AU \citep{Chiang1997}.
Model C has a constant disc aspect ratio $H/r$.
Observations of protostellar discs suggest that the  surface density parameter power law exponent $p$ is in the range of 0.5 to about 1, although there is considerable uncertainty  \citep{Williams2011}.  Model D has $p=0.5$, while other models have $p=1.0$.

\begin{table}
 \caption{Model Parameters}
 \label{tab:ModelParams}
 \begin{tabular}{lccccc}
  \hline
  Model & $H/r(r_{\rm in})$ & $s$ &  $p$ & $e_{\rm b}$ & $\alpha$ \\
  \hline
  A & 0.1   & 0.5  & 1.0   & 0.5 &  0.01 \\
  B & 0.05 & 0.5  &  1.0  &  0.5 & 0.01 \\
  C & 0.1   & 1.0  & 1.0   & 0.5 &  0.01 \\
  D & 0.1   & 0.5  & 0.5   & 0.5 &  0.01\\
  E & 0.1  &  0.5  & 1.0  &  0.75 & 0.01 \\
  F & 0.1  &  0.5  & 1.0  &  0.5 & 0.05 \\
    \hline
 \end{tabular}
\end{table}

\subsection{Tilt Evolution Rates} \label{sec:tiltevol}

For each disc model in Table \ref{tab:ModelParams} with outer disc edge $r_{\rm out}$
set to $50 a_{\rm b}$, we compute sequences of modes with various disc inner radii starting with large values of $r_{\rm in}$.
To calculate the longest lived modes, we determine the values  of $\ell_y(r_{\rm in})$ and $\omega$ 
for the large value of $r_{\rm in} = 49  a_{\rm b}$ using the nearly rigid tilt approximation of Section \ref{sec:expansion}.
At such radii, the tidal field is weak and the approximation holds with high accuracy.
We use those values as initial guesses for the shooting method with $r_{\rm in} = 49  a_{\rm b}$.
The shooting method converges rapidly and determines the numerically calculated eigenmode.
We then apply the numerically determined values of  $\ell_y(r_{\rm in})$ and $\omega$ obtained by the shooting method
iteratively as initial guesses for successively smaller values of $r_{\rm in}$, where the approximations made in Section \ref{sec:expansion} may not hold with good accuracy. 

\begin{figure}
\centering
\includegraphics[width=16.0cm]{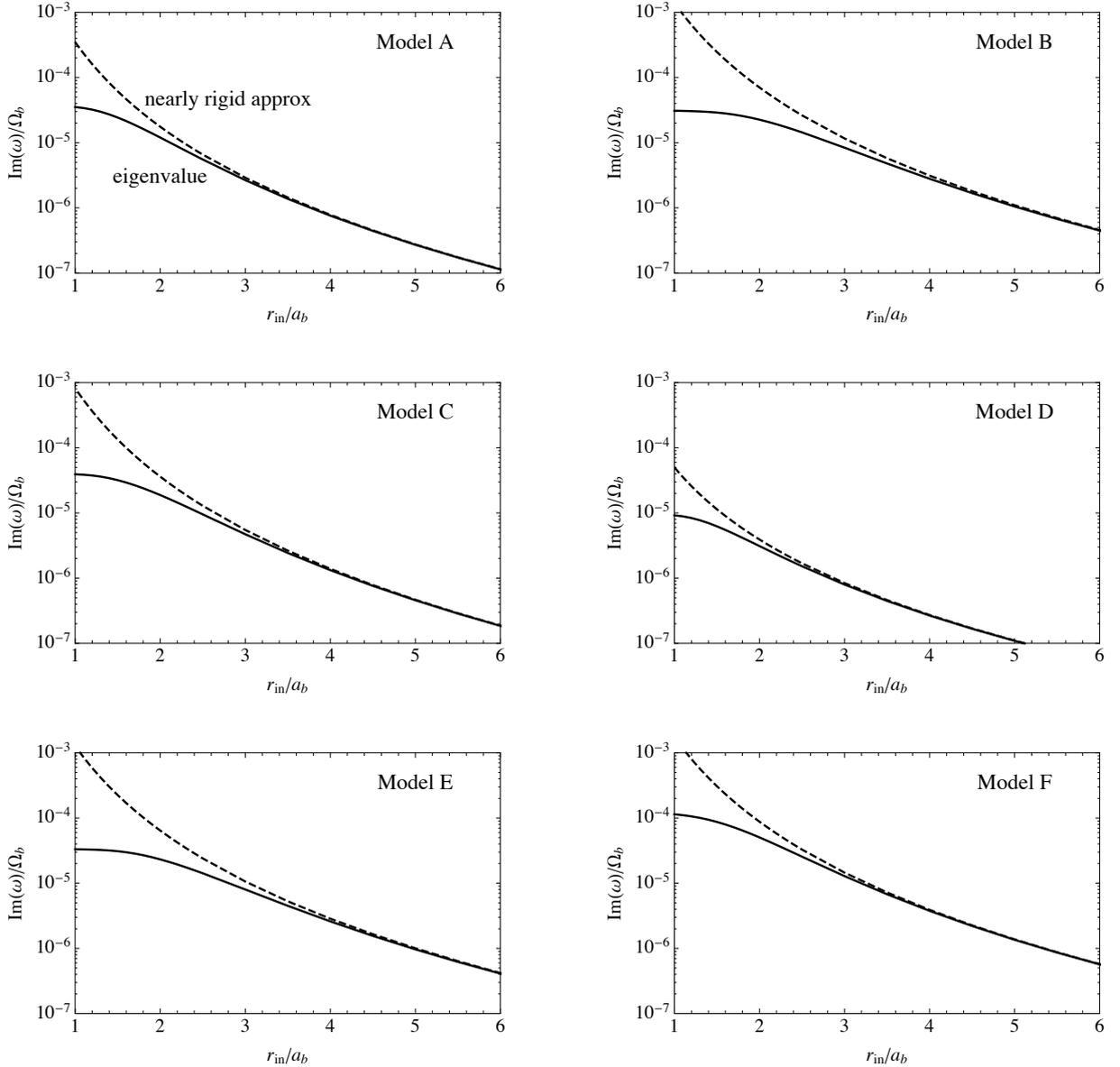}
\caption{Semi-log plots of the rates of change of tilt towards the polar direction as a function of the disc inner radius 
for the various models in Table \ref{tab:ModelParams}. The dashed line
is based on the nearly rigid tilt mode approximation, Equation (\ref{om2}), while the solid line is based on a numerically determined
eigenvalue.}
\label{fig:gr}
\end{figure}

Figure~\ref{fig:gr} plots the damping rates of tilt departures from the polar direction as a function
of disc inner radius for the various models listed in Table \ref{tab:ModelParams}. The numerically determined
damping rates are plotted as solid lines, while the damping rates determined by the nearly rigid tilt mode
approximation are plotted by the dashed lines. For larger disc inner radii, $r_{\rm in} \ga 3 a_{\rm b}$, the two sets
of lines merge, as is expected because the tidal effects of the binary on the disc are weak and consequently
the disc experiences little warping. At smaller   values of $r_{\rm in}$ the two sets of curves depart substantially,
indicative of significant warping which suggests that the linear equations may not be valid. We return to this point in Section \ref{sec:warping}.
In general, substantial evolution towards polar alignment occurs for all models over the course of a typical protostellar disc lifetime of a few million
years for  $r_{\rm in} \la 2 a_{\rm b}$ and binary orbital periods of the order of 100 years or less.

The results plotted in Figure~\ref{fig:gr}  across models qualitatively follow simple expectations.
Model B generally has a more rapid damping rate than Model A because the disc is cooler and therefore
experiences stronger warping because of the weaker pressure communication across the disc (see Equation  (\ref{td1})), as is discussed further
in Section \ref{sec:breaking}.  Model C has a lower temperature at larger radii than Model A and consequently experiences
stronger warping for a similar reason, resulting in a generally stronger damping. 
Model D has a weaker concentration of mass towards smaller radii
than Model A and consequently a larger fraction of the disc mass experiences the  weaker tidal
effects at larger radii, resulting in weaker damping. Model E has a higher binary eccentricity than Model A,
resulting in a more rapid disc precession and therefore a more rapid damping rate (see Equation (\ref{td2})).
Model F has a larger disc viscosity than Model A, resulting in a more rapid damping rate (see Equation  (\ref{td1})).

\begin{figure}
\centering
\includegraphics[width=12.0cm]{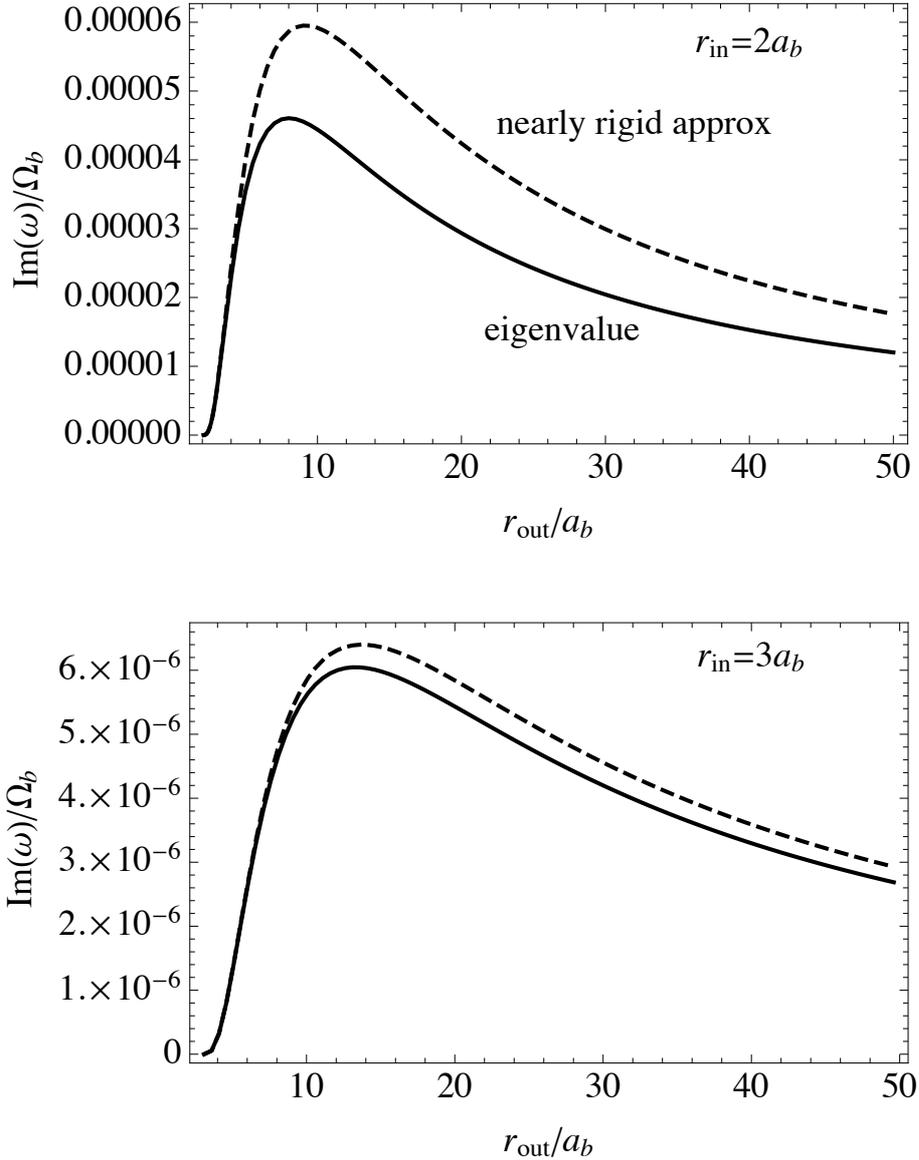}
\caption{Linear plots of the rates of change of tilt towards the polar direction as a function of the disc outer radius 
for Model A in Table \ref{tab:ModelParams}. The dashed line
is based on the nearly rigid tilt mode approximation, Equation (\ref{om2}), while the solid line is based on a numerically determined
eigenvalue. The upper (lower) panel has $r_{\rm in}=2 a_{\rm b}$ ($r_{\rm in}=3 a_{\rm b}$).
}
\label{fig:grout}
\end{figure}

Figure~\ref{fig:grout} plots the tilt damping rates of disc Model A towards polar alignment for a fixed  disc inner radius 
$r_{\rm in}=2 a_{\rm b}$ (upper panel) and $r_{\rm in}=3 a_{\rm b}$ (lower panel) and various disc outer radii.
These modes are  determined by an analogous procedure used for the determination of the modes in Figure~\ref{fig:gr}.
We apply as initial guesses for the shooting method  the values of $\ell_y(r_{\rm in})$ and $\omega$ 
by using the nearly rigid tilt approximation  described in Section \ref{sec:expansion}
at $r_{\rm out}=r_{\rm in}+0.1 a_{\rm b}$.  The disc is narrow and consequently the level of warping is expected to be small.  We then apply the numerically determined solutions  $\ell_y(r_{\rm in})$ and $\omega$ obtained by the shooting method
iteratively as initial guesses for successively larger values of $r_{\rm out}$, where the approximations made in Section \ref{sec:expansion} may not hold with good accuracy. 

For smaller disc outer radii,
the nearly rigid tilt mode approximation agrees well with the numerically determined eigenvalues because the disc
is somewhat narrow and therefore good radial communication is possible through pressure induced bending waves.
Consequently, there is little warping, even though the tidal forces are relatively strong. The agreement is better
for the lower plot with the larger disc inner radius where the tidal forces are weaker.

Note that the damping rate in Figure~\ref{fig:grout} has a peak value for $r_{\rm out} \sim 10 a_{\rm b}- 15  a_{\rm b}$
and is small for narrow and broad discs.  To qualitatively understand this behaviour, consider
Equation (\ref{om2}) for the tilt evolution rate $\omega^{(2)}$ in the nearly rigid tilt approximation.
The $\bm{G^{(1)}}$ term is related to the level of warping in the disc. $\omega^{(2)}$ is then
roughly related the the square of an average level of warping in the disc.
For a very narrow disc,
the level of warping is very small and therefore the tilt evolution rate is very small. This small level of warping
occurs because the radial communication timescale through pressure induced bending waves is very short.
For a very broad disc,  the average level of warping across the disc typically decreases
with increasing disc outer radius because the outer parts of the disc experience weak torquing by the binary.
Therefore for narrow discs, the tilt evolution rate increases with disc outer radius, while for very broad discs the tilt evolution
rate decreases, and there is a peak at intermediate values of disc width.

\subsection{Level of Warping} \label{sec:warping}

The tilt evolution Equations (\ref{lubow1}) and  (\ref{lubow2}) are linear and derived under the assumption
that the level of warping $r | d \bm{\ell}/dr|$ is in some sense small.  Since these equations were derived
using Eulerian perturbations, it is not clear from their derivation where the breakdown occurs.
\cite{Ogilvie2006} showed that nonlinear effects cause a significant deviation in the form of the bending wave
when  $r | d \bm{\ell}/dr| \sim (H/r)^{1/2}$ where the wave induced horizontal motions are sonic.  
For Model A, this implies that linearity  requires that $r | d \bm{\ell}/dr| \la 0.3$. 
 
\begin{figure}
\centering
\includegraphics[width=12.0cm]{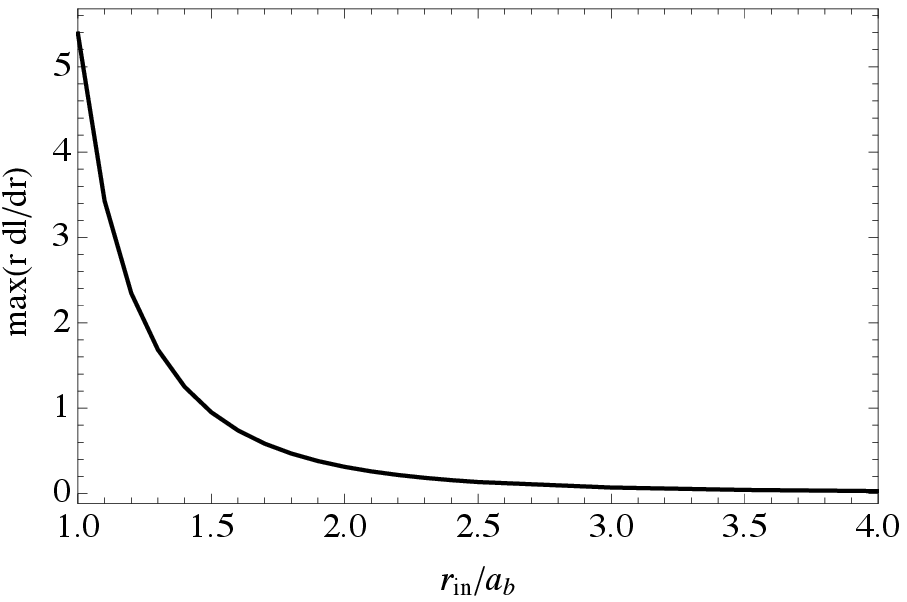}
\caption{ Maximum level of warping  $r | d \bm{\ell} /dr|$ across various discs as a function of their inner radii for Model A with fixed  $r_{\rm out}=50 a_{\rm b}$.
}
\label{fig:maxdldr}
\end{figure}

\begin{figure}
\centering
\includegraphics[width=12.0cm]{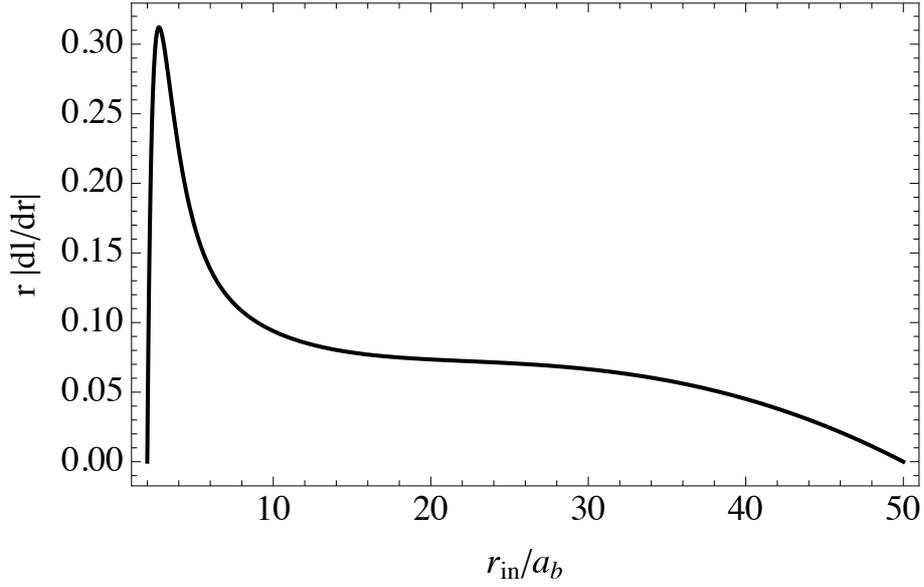}
\caption{ Level of warping  $r | d \bm{\ell}/dr|$ as a function of disc radius for the disc Model A with  $r_{\rm in}=2 a_{\rm b}$ and
$r_{\rm out}=50 a_{\rm b}$.
}
\label{fig:dldr}
\end{figure}

 Figure~\ref{fig:maxdldr} plots the level of maximum warping across various discs as a function 
of their inner radii for Model A with fixed $r_{\rm out}=50 a_{\rm b}$  which applies to the upper left panel of 
Figure~\ref{fig:gr}.
For disc inner radii smaller than about $2 a_{\rm b}$
this linearity requirement appears to be violated. Figure~\ref{fig:dldr} shows that the violation
occurs close to the disc inner radius where tidal forces are strong. Note that $d\bm{\ell}/dr=0$ at $r=r_{\rm in}$
and $r=r_{\rm out}$ due to boundary condition Equation (\ref{GBC}).

For  smaller $r_{\rm in}$, the maximum level of warping rises steeply with decreasing $r_{\rm in}$, $max(r | d \bm{\ell}/dr|) \sim r^{-4}$,
as determined by a rough fit to the numerically determined values. At smaller $r$, 
 $max(r | d \bm{\ell}/dr|) \sim r^{-4}$ also for Models C, D, and F in Figure~\ref{fig:gr}. 
For Models B and E, the level of warping increases
 more strongly with decreasing $r_{\rm in}$,  $max(r | d \bm{\ell}/dr|) \sim r^{-8}$.
 This rapid increase of warping at smaller $r_{\rm in}$ suggests that nonlinearity may be important
 even if a small fraction of the disc mass lies at such radii.

\begin{figure}
\centering
\includegraphics[width=12.0cm]{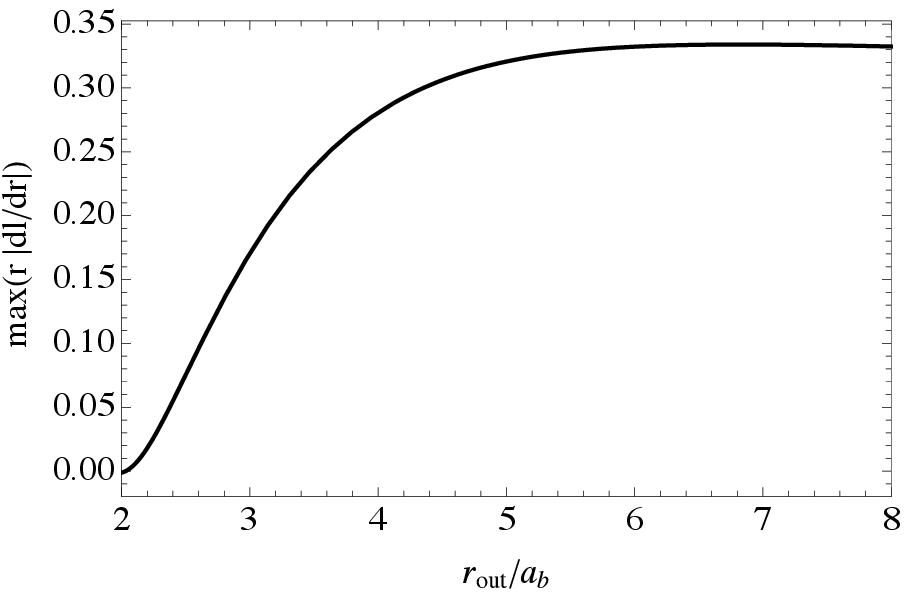}
\caption{ Maximum level of warping  $r | d \ell/dr|$ across various discs as a function of the disc outer radii for Model A with fixed $r_{\rm in}=2 a_{\rm b}$.
}
\label{fig:maxdldrro}
\end{figure}

 Figure~\ref{fig:maxdldrro} plots the level of maximum warping across various discs as a function 
of their outer radii for Model A with fixed $r_{\rm in}=2 a_{\rm b}$, which applies to the upper panel of 
Figure~\ref{fig:grout}.  
 In this case, the level of warping is small
for narrow discs where good radial communication is possible through pressure induced bending waves.
For discs with larger disc outer radii, the maximum level of  warping increases because the sonic communication timescale
across the disc increases. However, the maximum level of warping flattens at $r \ga 5 a_{\rm b}$ because
the outer parts of the disc experience weak tidal forcing.

\subsection{Approximate Criteria For Strong Warping/Breaking}
\label{sec:breaking}

In this section we consider an approximate criterion for when a broad circumbinary
disc ($r_{\rm in} \ll r_{\rm out}$) undergoes strong warping.
Strong warping could result in the disc breaking up radially into distinct annuli \citep[e.g.,][]{NK2012}.
 Estimates of breaking radii in the 
wave--like regime also have been made by \cite{Nixonetal2013} for circumbinary discs and \cite{Nealon2015} for discs around black holes.
The disc maintains radial communication via pressure induced bending waves that propagate at speed $c_{\rm s}/2$
for gas sound speed $c_{\rm s}$ \citep{Papaloizou1995, Lubowetal2002}.
This radial communication timescale is then approximately
\begin{equation}
t_{\rm c} \approx \frac{4}{ (2+ s) \Omega_{\rm b} h_{\rm out}} \left(\frac{r_{\rm out}}{a_{\rm b}} \right)^{3/2},
\label{tc}
\end{equation}
where $h_{\rm out}$ is the disc aspect ratio at the disc outer edge.

As we find in  Appendix A, in the quadrupole approximation for
the binary potential, the nodal precession rate for a test particle can generally be written as 
\begin{equation}
\omega_{\rm n} (r)=  k \, \left( \frac{a_{\rm b}}{r} \right)^{7/2}  \Omega_{\rm b},
\label{omnr}
\end{equation}
where $k$ is constant in $r$ that depends on the binary mass ratio and eccentricity and disc relative orientation.
We determine global precession rate of a disc by taking its angular momentum average of $\omega_{\rm n}$,  as indicated
by Equation (\ref{om1}). 
As in Section \ref{sec:models}, we consider  discs whose density varies in $r$ as $\Sigma \propto 1/r^p$, for some constant $0.5 \le p \le 1$. We take the precession timescale $t_{\rm p}$ to be the inverse
of the global precession rate and obtain for $r_{\rm in} \ll r_{\rm out}$
\begin{equation}
t_{\rm p}=  \frac{2 (1+p) r_{\rm in}^{1+p} r_{\rm out}^{5/2-p} }{|k|(5 -2 p) a_{\rm b}^{7/2} \Omega_{\rm b}}.
\label{tp}
\end{equation}

On timescale $t_{\rm p}$, the disc would under differential precession resulting in strong warping  or breaking in the absence
of radial communication.
Breaking can occur for a disc whose inner radius is sufficiently small that
that $t_{\rm c} > t_{\rm p}$.
By equating the precession timescale $t_{\rm p}$ to the sonic communication timescale $t_{\rm c}$, the breaking inner disc radius in the wave-like regime is then estimated as
\begin{equation}
r_{\rm break}\approx\left( 
\frac{2 (5-2 p) |k|  }{h_{\rm out} (2+s+2p + p s)}   \left( \frac{a_{\rm b}}{r_{\rm out} } \right) ^{1-p} 
 \right)^\frac{1}{1+p}a_{\rm b}
 \label{rbr}
\end{equation}
\citep[see also][for the viscous regime where $\alpha > H/r$]{Nixonetal2013}.
Notice that in the case $p=1$ as in Model A, we have that $r_{\rm break}$ is independent of the disc outer radius $r_{\rm out}$ for fixed $h_{\rm out}$.

For a nearly coplanar disc with $p=1$ and $s=0.5$ that orbits a circular binary with equal mass component stars, we have from Equation (\ref{nodalop}) 
that $k=3/16$ and therefore $r_{\rm break} \approx 1.5   (0.1/h_{\rm out})^{1/2} a_{\rm b}$.
For $p=0.5$, $s=0.5$, and $r_{\rm out}=50 a_{\rm b}$, we have that
$r_{\rm break} \approx 0.7 (0.1/h_{\rm out})^{2/3} a_{\rm b}$.  For $H/r \sim 0.1$ as can occur in the protostellar disc case,
breaking does not appear to be likely because the breaking radius is typically smaller than the tidal inner truncation radius
of a circumbinary disc  determined by \cite{Artymowicz1994}(hereafter AL94).

For a nearly polar disc that follows Model A with equal mass component stars  and $e_b=0.5$, we have from Equation (\ref{om1})
that $k \approx 0.30$ and therefore 
$r_{\rm break} \approx 1.9 (0.1/h_{\rm out})^{1/2} a_{\rm b}$.
In Figure~\ref{fig:maxdldr} the breaking radius is expected to occur where the maximum level of warping is of order unity.
To apply this relation to Figure~\ref{fig:maxdldr},  we use the fact that $h_{\rm out }=0.1 (50 a_b/r_{\rm break})^{1/4}$  for Model A.
We then predict that 
 $r_{\rm break} \simeq 1.2 a_{\rm b}$, where the maximum level of warping is $\sim 2.5$ in the figure,
in rough agreement with expectation that the level of warping is of order unity.
For $p=0.5$, $s=0.5$, and $r_{\rm out}=50 a_{\rm b}$ we have that
$ r_{\rm break} \approx 0.9 (0.1/h_{\rm out})^{2/3} a_{\rm b}$.


\subsection{Contributions to Tilt Evolution Rate}
\label{sec:omcont}

\begin{figure}
\centering
\includegraphics[width=8.0cm]{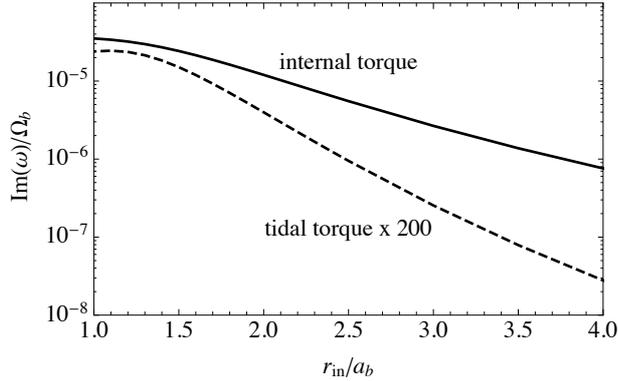}
\caption{Semi-log plot of the contributions to the tilt evolution rate towards polar orientation as a function of disc inner radius $r_{\rm in}$
for fixed disc outer radius $r_{\rm out}= 50 a_{\rm b}$ with disc Model A. The solid line is the contribution due to the disc internal torque 
$\bm{G}$ term on the RHS of Equation (\ref{Imom}) and the dashed line
is the contribution due to the binary torque $\bm{\tau}$ term multiplied by 200.  }
\label{fig:omcont}
\end{figure}

There are two contributions to the rate of change of tilt towards polar, $Im(\omega)$ in Equation (\ref{Imom}). 
One contribution involves a tidal torque term, the first
term on the RHS of Equation (\ref{Imom}), and another contribution is due an internal torque $\bm{G}$ term,
the second term on the RHS. We showed in Section \ref{sec:higherorder} that the tidal torque
contribution should be much smaller than the internal torque contribution, since the former vanishes
to two orders in the nearly rigid tilt expansion based on the weakness of the tidal field, while the latter does not vanish at this order. 
In Figure~\ref{fig:omcont} we plot the contributions involving these two terms to the tilt evolution
rate $Im(\omega)$ using the results of the numerically determined modes for disc Model A. 
The plot is for the same case shown in the upper left panel of Figure~\ref{fig:gr}.
The contributions from the tidal term rise for smaller disc inner radii where the tidal field becomes stronger.
As expected, its contributions are much smaller than those of the internal torque term, but are not vanishingly small.

\section{Inner Truncation of a Polar Disc}
\label{sec:truncation}

\subsection{Role of Resonances}
The inner tidal truncation of coplanar circumbinary discs involving eccentric orbit binaries is likely
dominated by the effects of eccentric Lindblad resonances (AL94).
At such resonances, waves are launched that carry angular momentum. As the waves damp,
the angular momentum carried by the waves is transferred to the disc \citep{Goldreich1979}.
The gap opening condition is that the outward Lindblad resonant torque exceeds the
inward viscous torque. There are infinitely many Lindblad resonances at various radii in a circumbinary
disc. The outermost (weakest) Lindblad resonance in a circumbinary disc that satisfies the gap opening
condition is the one that controls the circumbinary disc inner truncation.
 In the case that the waves damp rapidly near the resonance,
this outermost resonance location determines the circumbinary disc inner radius.  

AL94 determined  circumbinary disc  inner radii by this procedure for discs
that are coplanar and rotate in the same sense as the central binary. 
For tilted circumbinary discs with respect to the orbital plane of the binary, 
the Lindblad resonance torques are weaker than the coplanar case, resulting in a
 decreasing central hole size with tilt
\citep{Nixon2015,Miranda2015}.
The reasons for the decrease are that the disc can lie farther away from the binary in the noncoplanar case
and also because the relative speeds of the disc and binary increases with tilt.

\subsection{Torque Equation}
We determine the resonant torque exerted on a polar circumbinary disc due to an eccentric binary. 
We closely follow the approach of  \cite{Nixon2015} that was applied to counter rotating discs.
This approach has advantages over previous methods 
\citep[e.g.,][]{Goldreich1980,Artymowicz1994,Miranda2015}
that perform expansions in the binary eccentricity that is not generally small for binaries. 
The calculation of the resonant torque involves a Fourier transform of the potential in time.
The expansion was used  because there does not exist an explicit analytic expression for instantaneous radius of the binary orbit in terms of the binary mean anomaly (time). 
The approach we apply, based on \cite{Moorhead2008}, determines the resonant torque directly numerically without
performing an expansion in eccentricity. The advantage comes about because the instantaneous radius of the binary can
be expressed explicitly in terms of its eccentric anomaly that is used as a variable, instead of mean anomaly that was used previously. The Fourier transform is expressed in terms of the eccentric anomaly, rather than time. 

 We adopt a cylindrical coordinate system $(r, \theta, z)$ whose origin is at the center of mass and so 
the $z=0$ plane coincides with the disc midplane.  Following \cite{Goldreich1979}, we decompose the potential as
\begin{equation}
\Phi(r, \theta, t) = \sum_{l, m} \Phi_{l, m}(r) \cos{(m \theta - l \Omega_{\rm b} t)}\,,
\label{PhiGT}
\end{equation}
where $l$  (not to be confused with disc tilt vector $\bm{\ell}$) and $m$ range over integers (zero, and positive), and $\Omega_{\rm b}>0$. The binary potential has contributions from the primary star of mass $M_1$ and the secondary star of mass $M_2$. We determine the potential components $\Phi_{l, m}$ for $l>0$ and  $m \ge 0$ 
by  
\begin{equation}
 \Phi_{\ell, m}(r) = \Phi^{(1)}_{l, m}(r)+ \Phi^{(2)}_{l, m}(r),
 \label{Plm}
\end{equation}
where the superscripts 1 and 2 denote the primary and secondary objects. We invert Equation~(\ref{PhiGT}) by using the eccentric anomaly of the binary orbit $\zeta$ as a variable of integration in place of the mean anomaly, $\Omega_{\rm b} t$. 
 For a disc that is tilted  with respect to the binary orbital plane
 by angle $\psi$  about the axis
defined by $ \bm{e}_{\rm b}  \times \bm{J}_{\rm b}$, we then have for $i=1,2$ 
\begin{equation}
\Phi^{(i)}_{l, m}(r) = -\frac{G M_i}{2 \pi^2 a_{\rm b}} \left(1- \frac{\delta_{m,0}}{2} \right) \int_0^{2 \pi}d \theta  \int_0^{2 \pi} d \zeta\,\, \frac{\left(1- e_{\rm b} \cos{\zeta}\right)\cos{( m \theta -\ell (\zeta - e_{\rm b} \sin{\zeta}) )}}{\sqrt{v^2 + x_i^2 (1-e_{\rm b}  \cos{\zeta})^2- 2 v x_i g(\theta,\zeta)}},
\label{Plmi}
\end{equation}
where
\begin{eqnarray}
g(\theta,\zeta) = \cos{(\psi)} (\cos{\zeta} -e_{\rm b}) \cos{(\theta)}+\sqrt{1-e_{\rm b}^2} \sin{(\zeta)} \sin{(\theta)},
\end{eqnarray}
$x_1 = - M_2/(M_1+M_2)$ and $x_2=M_1/(M_1+M_2)$ and $v=r/a_{\rm b}$. For a polar disc $\psi=\pi/2$, while for a coplanar disc  $\psi=0$.

The outer Lindblad resonances in a Keplerian circumbinary disc  occur
where
\begin{equation}
 \Omega(r_{l,m}) = \frac{l \Omega_{\rm b}} {m+1}.
\label{Omr}
\end{equation}

\subsection{Results}

The eccentric outer Lindblad resonance with $(l, m)=(1,2)$ that occurs where $\Omega(r)= \Omega_{\rm b}/3$ (via Equation (\ref{Omr})) plays an important role in the inner truncation of circumbinary discs of moderate viscosity and eccentricity
(AL94). It is the outermost first order eccentric Lindblad resonance. For such a resonance, the torque varies quadratically in $e_{\rm b}$ for $e_{\rm b} \ll 1$. Resonances that occur at larger radii involve torques that vary as higher powers of  $e_{\rm b}$ and are weaker for $e_{\rm b} \ll 1$.
The outer Lindblad resonance with $(l, m)=(2,2)$ that occurs where $\Omega(r)= 2 \Omega_{\rm b}/3$ is closer to the binary. It is the outermost noneccentric  Lindblad resonance for an equal mass binary.  Since it is closer
to the binary than the $(1,2)$ resonance, its effects on disc truncation are expected to be stronger.
 
\begin{figure}
\centering
\includegraphics[width=14.0cm]{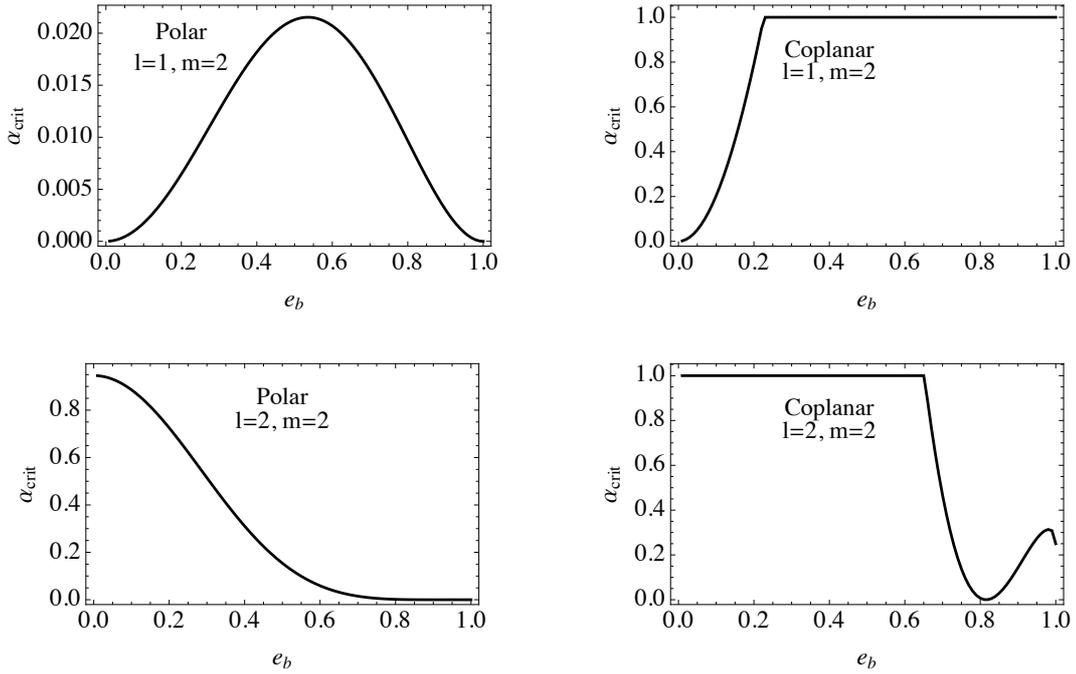}
\caption{ Critical values of viscosity parameter $\alpha$ as a function of  binary eccentricity for the $(l,m)=
(1, 2)$ and $(2, 2)$ Lindblad resonances
that occur at $\Omega(r)=\Omega_{\rm b}/3$ and $\Omega(r)=2 \Omega_{\rm b}/3$, respectively, for equal mass binary components with disc aspect ratio $H/r=0.1$ at the resonance.
}
\label{fig:alcrit}
\end{figure}

Using Equation (16) of AL94 that balances viscous with resonant torques, we determine the critical viscosities $\alpha_{\rm crit}$ by 
\begin{equation}
\alpha_{\rm crit}= min\left(\pi m \left( \frac{\Phi_{\ell, m} }{G M/a_{\rm b}} \,  \frac{(m+1)^{1/6}(3 m+1)}{ 3 l^{2/3}} \frac{r}{H}\right)^2, 1 \right),
\label{alphacr}
\end{equation}
where $\Phi_{\ell, m}$ is given by Equations (\ref{Plm}) and (\ref{Plmi}).
For $\alpha < \alpha_{\rm crit}$ the resonant
torque is able to open a gap in the disc.
  The $min$ is taken in Equation (\ref{alphacr})
because values of $\alpha$ greater than unity are unphysical.
The application of these results to tilted discs in the wavelike regime that we consider in this paper 
further requires that $\alpha < H/r$.

We apply Equation (\ref{alphacr})
to  the $(1, 2)$ and $(2, 2)$ resonances.
Figure~\ref{fig:alcrit} plots the $\alpha_{\rm crit}$ values for this resonance
as a function of binary eccentricity for disc aspect ratio $H/r=0.1$ and equal mass binary component stars in both the polar and coplanar cases.
The $\alpha_{\rm crit}$  values scale inversely with $(H/r)^2$, and so a thinner disc would have higher values of $\alpha_{\rm crit} < 1$.
For $(l, m)=(1,2)$ in both  the coplanar and polar  cases, $\alpha_{\rm crit}$ goes to zero as $e_{\rm b}$ goes to zero
because the resonant torque vanishes at $e_{\rm b}=0$ for this eccentric Lindblad resonance.
For $(l, m)=(2,2)$ in both  the coplanar and polar  cases, $\alpha_{\rm crit}$ does  not go to zero as $e_{\rm b}$ goes to zero.

In the coplanar case, the $(l,m)=(1,2)$ resonance can truncate a disc for even fairly small values of binary eccentricity.
Any value of $\alpha < 1$ can result in gap opening for $e_{\rm b} > 0.25$.
Tidal truncation of a polar disc at this resonance is harder to achieve, but is possible for $\alpha \sim 0.01$ and moderate values
of binary eccentricity $e_{\rm b} \sim 0.5$. This resonance occurs at $r \simeq 2.08 a_{\rm b}$.
We see from Figure~\ref{fig:maxdldr} that at this radius, a nearly polar disc marginally satisfies the conditions for the assumed
linearity of the tilt evolution equations used in this paper.

In coplanar case, the $(l, m)=(2, 2)$ resonance can generally
truncate the disc, for any value of $\alpha < 1$, for $e_{\rm b} \la 0.7.$ $\alpha_{\rm crit}$ vanishes near $e_b=0.8$
because the potential component $\Phi_{2, 2}$ vanishes near there. 
In polar case, $\alpha_{\rm crit}$ for the $(l, m)=(2, 2)$ resonance declines with increasing binary eccentricity.

In both polar cases, $\alpha_{\rm crit}$
decreases at larger eccentricity as the binary orbit becomes more elongated.
In the extreme polar case of $e_{\rm b}=1$, the binary orbit takes the form of a line that lies perpendicular
to the disc plane and provides an axisymmetric potential in the disc plane.
Thus, the torque must vanish in that case, as seen in the left panels of Figure~\ref{fig:alcrit}.
By this argument, it follows that all resonant tidal torques vanish for $e_{\rm b}=1$.
Consequently, we expect that a highly eccentric binary is unlikely to be able to efficiently tidally truncate
a polar disc. As as result, the polar circumbinary disc will likely undergo strong warping and possibly breaking at smaller radii.

\section{Unstable  Perpendicular Configuration}
\label{sec:perp}
Stationary test particle orbits in the inertial frame around an eccentric binary exist for coplanar and polar orbit orientations.
They are related to the stable circumbinary disc configuration for the coplanar case \citep{Facchinietal2013, Foucart2014}
and also for the polar state as we saw in Section \ref{sec:higherorder}. In the coplanar and polar cases, discs undergo nodal precession 
for the small departures from a coplanar/polar state. They evolve towards their final states due to the effects of dissipation.
As pointed out by \cite{Farago2010}, there is an orthogonal direction of particle angular momentum, along $\bm{J}_{\rm b} \times \bm{e}_{\rm b}$, for which test particle orbits are stationary in the inertial frame, but are unstable. To investigate the disc evolution in that case, we proceed along the lines of the analysis
in Section \ref{sec:model}.

We take the $x$-axis to be along $\bm{J}_{\rm b}$ and the $z$-axis to be along  $\bm{J}_{\rm b} \times \bm{e}_{\rm b}$ 
and apply equations (2.16) and (2.18) in \cite{Farago2010}.  For the torque defined in Equation (\ref{Ttor}) we then have
\begin{equation}
a(r) =5e_{\rm b}^2 \, \omega_{\rm p}(r)
\label{tauxp}
\end{equation}
and
\begin{equation}
b(r)=(1-  e_{\rm b}^2) \, \omega_{\rm p}(r)  ,
\label{tauyp}
\end{equation}
with $\omega_{\rm p}$ again given by Equation (\ref{omegap}).
Note that we apply
a different coordinate  system in which
our $(x,y,z)$ coordinates corresponds to their  $(z, x, y)$ coordinates. The binary orbit lies in the $y-z$ plane.
In the rigid tilt expansion in analogy with Equation (\ref{om1}) for the polar case, we have
\begin{equation}
\omega^{(1)} = \pm i \frac{3\sqrt{5}}{4}  e_{\rm b} \sqrt{1- e_{\rm b}^2} \frac{M_1 M_2}{M^2} \Omega_b \left<\left(\frac{a_{\rm b}}{r} \right)^{7/2} \right>.
\label{om1u}
\end{equation}
In this case, $\omega^{(1)}$ is imaginary an therefore corresponds to an unstable tilt evolution in lowest order.
Independent of the results at next order, the disc generally undergoes growing deviations of tilt away from the perpendicular state.

For this perpendicular configuration, the first term on the RHS of Equation (\ref{lam})   is nonzero in first order and dominates over the second term. This behaviour is quite different from the polar case where this term is zero to second
order.

\section{Energetics}
\label{sec:energetics}

As noted in the Introduction, a circumbinary test particle orbit of fixed radius that is nearly
coplanar to a binary and subject to some form of dissipation would be expected to become coplanar, based 
on the idea that the coplanar orbit has the least energy.  
However, the polar case is at a global energy
maximum for particle orbits of fixed radius over all possible orbital orientations. This can be seen from the Hamiltonian given by Equation (2.13) of \cite{Farago2010} for which the polar case has tilt $i_2=\pi/2$ and longitude of ascending node $\Omega_2=\pi/2$ in their notation. The coplanar case ($i_2= 0$) is at a global energy minimum and the perpendicular case of Section \ref{sec:perp} ($i_2=\pi/2, \, \Omega_2=0$)
is intermediate.

This somewhat counter intuitive result that polar orbits are stable at energy maxima has been examined
in the context of a possible polar ring around Neptune by \cite{Dobrovolskis1989}.
They describe a somewhat simplified case of a viscous ring around a spheroid that is axisymmetric about the polar axis,
rather than the triaxial ellipsoid case with unequal axes that applies to an eccentric binary. However, the main physical point is relevant. Based on angular momentum conservation in the polar direction,
they show that such a nearly polar ring must contract (accrete) as it achieves a polar orientation. Therefore, the assumption of fixed radius
in comparing orbital energies is not correct. When the associated ring contraction is accounted for, there is a net energy
loss in evolving to the polar state.

\cite{Dobrovolskis1989} also point out that a viscous ring that orbits about an oblate spheroid
and is initial slightly tilted with respect to spheroid's  equator would be expected to evolve
to the plane of the equator. This configuration is analogous to that of a 
 circumbinary disc that is initially nearly coplanar. On the other hand,
consider evolution of a viscous ring about a prolate spheroid that is initial slightly tilted with respect to its equator. This configuration is analogous to that of a  circumbinary disc that is initially nearly polar.
 If the two cases have quadrupole moments of equal magnitude
but opposite sign, then the rings undergo nodal precession in opposite directions in the two cases.
But by Equation (\ref{td1}) applied to such systems, both cases should evolve to equatorial alignment at equal rates. 

Polar secular stability is made possible because slightly   tilted test particle orbits away from the polar state 
undergo precession and are therefore neutrally stable. Dissipational forces acting on this neutral state
are able to cause evolution towards the polar state.
The perpendicular state discussed in Section \ref{sec:perp} is unstable to slight departures from the polar state.
Its evolution corresponds to simple expectations of evolution from a local energy maximum.

\section{Summary}
\label{conc}

We have investigated the evolution of  a nearly polar low mass disc around an eccentric binary by means of linear theory. Evolution towards the 
polar state was
recently shown in SPH simulations by \cite{Martin2017}. 
We apply the secular binary potential in the quadrupole approximation developed by \cite{Farago2010}
to explore the secular evolution of a nearly polar disc by using the linear tilt evolution equations
of \cite{Lubow2000}. 
Unlike the circumbinary disc evolution in the circular orbit binary case,
in the eccentric binary case the disc undergoes secular tilt oscillations, as well as precession (see Figure \ref{ivrst}).

We have shown analytically in the limit of weak tidal forcing that a nearly polar circumbinary
disc evolves towards the polar state through the effects of dissipation caused by turbulent viscosity.  We have determined the tilt evolution
timescales by means of an analytic nearly rigid tilt expansion and by means of numerically determined
modes. The two methods agree well where the disc warping is small (see Figures \ref{fig:gr} and \ref{fig:grout}). 
For typical protostellar disc parameters and binary eccentricity $e_{\rm b} \sim 0.5$,
substantial tilt evolution can occur over a typical disc lifetime of a few million years
for equal mass binaries whose orbital periods are of order 100 years or less.
Estimates of evolution timescales for other parameters can be obtained using scaling relation
Equation (\ref{td2}).
The tilt evolution timescales fall rapidly with the decreasing radius of the disc inner edge.
For disc inner radii of a few times the binary semi-major axis, the warping becomes
strong enough that the linearity assumption begins to break down (see Figures \ref{fig:maxdldr} and \ref{fig:dldr}).

Disc tidal truncation by means of resonance torques involving eccentric binaries  is much less
effective for polar discs than for coplanar discs (see Figure \ref{fig:alcrit}). For a binary with an eccentricity of unity,
no resonant forcing of a polar disc occurs because the binary provides an axisymmetric secular potential in the disc plane.
The results suggest then that polar disc material in Keplerian orbits may sometimes reach to small
enough disc radii that nonlinear effects may play an important role.

For fixed orbital radius, polar orbits are at an energy maximum, while coplanar
orbits are at an energy minimum.
In Section \ref{sec:energetics}, we describe how disc evolution to a polar state is energetically possible.

 In this analysis we have considered only low mass discs, whose angular momenta are small
compared to that of their central binary. For discs of larger mass, the gravitational and accretion torques
exerted by the disc on the binary can have an important influence on the binary orbital evolution.
As noted in \cite{Martin2017}, such effects can reduce the binary eccentricity that may in turn suppress
disc polar alignment. 

The results show that circumbinary discs can evolve to polar states, such as found in observations of binary 99 Herculis
 \citep{Kennedy2012}. Circumbinary planets might then form in such discs and reside on polar orbits.
 The long term stability of planets on such orbits needs to be understood.
 The analytic model of \cite{Farago2010} suggests that polar orbits of test particles are stable.
 However, this model is based on the quadrupole approximation for the binary potential. \cite{Doolin2011} carried out direct test particle orbit integrations. Their results
 revealed that polar orbits are stable over timescales of $5 \times 10^4$ binary orbits (the maximum timescale of their simulations) for order unity mass ratio binaries
 with eccentricities of a tenth or more,
 if the orbital radius is greater than a few times the binary semi-major axis.



\bibliographystyle{mnras} 
\bibliography{main}

\appendix
\section{Precession Rates}

We consider a massless particle orbiting a central binary with components of mass $M_1$ and $M_2$ and total mass $M=M_1+M_2$, semi-major axis $a_{\rm b}$, the angular frequency $\Omega_{\rm b}=\sqrt{GM/a_{\rm b}^3}$,
and eccentricity $e_{\rm b}$.  The particle is taken to
be on a circular orbit around the binary. We work in a frame in which the angular momentum of the binary is along the $z$--axis and the eccentricity vector of the binary is along the $x$--axis. The position of the particle relative to the barycentre of the binary is $\bm{r}_{\rm p}=(x,y,z)$. The particle is at a distance $r_{\rm p}=\sqrt{x^2+y^2+z^2}$.  The  instantaneous
difference in positions of the binary components is denoted by  $\bm{r}_{\rm b}$.

Assuming that $r_{\rm b}\ll r_{\rm p}$,  the gravitational potential due to the binary is given by 
\begin{equation}
\Phi=\Phi_{\rm K}+\Phi_{\rm p},
\label{Phi}
\end{equation}
where
\begin{align}
\Phi_{\rm K}=& -\frac{G M}{ r_{\rm p}} 
 \label{PhiK}
\end{align}
is the Keplerian potential 
and
\begin{align}
\Phi_{\rm p}=& -\frac{G \beta}{2 r_{\rm p}^3}\left[3\frac{(\bm{r}_{\rm p}\cdot \bm{r}_{\rm b})^2}{r_{\rm p}^2} -r_{\rm b}^2\right] 
\end{align}
is the lowest order (quadrupole) perturbing potential
and the reduced mass is
\begin{equation}
\beta=\frac{M_1M_2}{M_1+M_2}.
\end{equation}

In order to determine the secular evolution, we need to average the perturbing potential
over the mean anomaly (time) for the orbits of the binary and the particle. 
We first consider the time average over the binary orbit. As shown in \cite{Farago2010}, 
\begin{equation}
\left< \bm{r}_{\rm b}\cdot \bm{r}_{\rm p} \right>^2=\frac{a_{\rm b}^2}{2}(x^2+y^2)+\frac{a_{\rm b}^2e_{\rm b}^2}{2}(4x^2-y^2)
\end{equation}
and
\begin{equation}
\left< r_{\rm b}^2 \right> = a_{\rm b}^2\left(1+\frac{3}{2}e_{\rm b}^2\right).
\end{equation}
The binary orbit time-averaged potential is then
\begin{align}
\left <\Phi_{\rm p} \right>=& 
 -\frac{a_{\rm b}^2 G \beta \left[(1+9e_{\rm b}^2)x^2+(1-6 e_{\rm b}^2)y^2-(2+3e_{\rm b}^2)z^2\right]}{4 (x^2+y^2+z^2)^{5/2}}.
\end{align}

We define a new coordinate system in which the $z'$ axis is along the angular momentum vector of the particle, 
\begin{equation}
\bm{\ell}_3=\bm{\ell}_{\rm p}=(\ell_x,\ell_y,\ell_z).
\end{equation}
The unit vector along the direction of angular momentum of the binary $\bm{\ell}_{\rm b}=(0,0,1)$ in the original coordinate system. The new $y'$ axis is in the direction
\begin{equation}
\bm{\ell}_2
=\frac{\bm{\ell}_{\rm b}\times \bm{\ell}_{\rm p}}{|\bm{\ell}_{\rm b}\times \bm{\ell}_{\rm p}|}
=\left(-\frac{\ell_y}{\sqrt{\ell_x^2+\ell_y^2}},\frac{\ell_x}{\sqrt{\ell_x^2+\ell_y^2}},0\right)
\end{equation}
 and the $x'$ axis is in the direction
\begin{equation}
\bm{\ell}_1=\frac{\bm{\ell}_2\times \bm{\ell}_3}{|\bm{\ell}_2\times \bm{\ell}_3|}.
\end{equation}
 We can convert the angular momentum of the particle to the new coordinate system with
\begin{equation}
\bm{r}_{\rm p}'=(x',y',z')=(\bm{r}_{\rm p}\cdot \bm{\ell}_1,\bm{r}_{\rm p}\cdot \bm{\ell}_2,\bm{r}_{\rm p}\cdot \bm{\ell}_3).
\end{equation}
We solve this equation to find $x$, $y$ and $z$ in terms of $x'$, $y'$ and $z'$.

We work in plane polar coordinates in the plane of the circular particle orbit and set $x'=r \cos \phi$, $y'=r \sin \phi$ and $z'=0$. 
In order to average the potential over the mean anomaly of the particle, we average over $\phi$ and obtain
\begin{align}
\left<\left<\Phi_{\rm p} \right>\right>=  \frac{G \beta a_{\rm b}^2  }{8 r^3} 
\left[3 e_{\rm b}^2 \left(4 \ell_x^2-\ell_y^2-1\right)+3
   \ell_x^2+3 \ell_y^2-2\right],
 \label{Phip}
\end{align}
where we have eliminated $\ell_z$ through the identity $\ell_z^2=1-\ell_x^2-\ell_y^2$.
We  denote the doubly averaged potential
as $\Psi= \left< \left<  \Phi_{\rm K}+\Phi_{\rm p}\right> \right>$.
The particle angular frequency is given by
\begin{equation}
\Omega^2=\frac{1}{r} \Psi'(r)
\end{equation}
and the epicyclic frequency
\begin{equation}
\kappa^2=\frac{3}{r}\Psi'(r)+\Psi''(r),
\end{equation}
where prime denotes differentiation.
Using Equations (\ref{Phi}), (\ref{PhiK}), and (\ref{Phip}),
we then find the apsidal precession rate is given by
\begin{align}
\omega_{\rm a}= &\, \Omega-\kappa \simeq \frac{\Omega^2-\kappa^2}{2\Omega}\notag \cr
\approx &-\frac{3}{8}\Omega_{\rm b}\left(\frac{\beta }{ M } \right)
\left(\frac{a_{\rm b}}{r}\right)^{7/2}  \left[3 e_{\rm b}^2 \left(4 \ell_x^2-\ell_y^2-1\right)+3
   \ell_x^2+3 \ell_y^2-2\right],
\end{align}
where  $\Omega = \sqrt{GM/r^3}$ is the Keplerian velocity.
For a particle orbit that is close to coplanar, so that $l_x=l_y=0$, the apsidal precession rate is given by
\begin{equation}
\omega_{\rm a}=\frac{3}{8}\Omega_{\rm b}\left(\frac{\beta }{ M } \right)
\left(\frac{a_{\rm b}}{r}\right)^{7/2}
\left(2 +3 e_{\rm b}^2 \right).
\label{op}
\end{equation}
For a particle orbit that is close to polar, such that $\ell_x=1$ and $\ell_y=0$, the apsidal precession rate is given by
\begin{equation}
\omega_{\rm a}=-\frac{3}{8}\Omega_{\rm b}\left(\frac{\beta }{ M } \right)
\left(\frac{a_{\rm b}}{r}\right)^{7/2}
\left(1+9 e_{\rm b}^2\right).
\label{opp}
\end{equation}

For a particle orbit that is nearly coplanar about a circular orbit binary, the nodal and apsidal precession rates are nearly
equal in magnitude, but opposite in sign. Therefore, from Equation (\ref{op}), 
the nodal precession rate about a circular orbit binary
in that case is given by
\begin{equation}
\omega_{\rm n}=-\frac{3}{4}\Omega_{\rm b}\frac{\beta }{ M } 
\left(\frac{a_{\rm b}}{r}\right)^{7/2},
\label{nodalop}
\end{equation}
which agrees with equation (2.24) of \cite{Farago2010} derived for a circular orbit binary with $i_2=0$ and $e_2=0$ in their notation.

\section{Evaluation  of Unit Vector product term }

\subsection{Constancy of Unit Vector Product Term }

We  define quantity
\begin{equation}
\chi(r) = Re(\ell_x(r) \ell_y^*(r))
\label{chi}
\end{equation}
and consider  its derivative in first order as 
\begin{equation}
\frac{d \chi^{(1)}}{dr} = Re\left(\frac{d \ell^{(1)}_x}{dr} \, \ell^{(0)*}_y + \ell_x^{(0)} \, \frac{d \ell^{(1)*}_{y}}{dr} \right).
\end{equation}
By multiplying  the $x$-component of Equation (\ref{l2a})  by $\ell^{(0)*}_y$ and multiplying  the $y$-component of Equation (\ref{l2a})  by $\ell^{(0)*}_x$
we obtain
\begin{equation}
\frac{d \chi^{(1)}}{dr} = \frac{4 \alpha \Omega}{{\cal I} r^3 \Omega^3} Re(G^{(1)}_{x} \, \ell^{(0)*}_y + G^{(1)}_{y}  \, \ell^{(0)*}_x). 
\end{equation}
By applying  Equations (\ref{om1}), (\ref{ly0}), and (\ref{G0}), we obtain that
\begin{equation}
Re(G^{(1)}_{x} \, \ell^{(0)*}_y) =  \left|\ell_x^{(0)} \right|^2 \frac{\left< B^{(1)} \right>}{\left< A^{(1)} \right>}  \int_{r_{\rm in}}^r \Sigma r'^3 \Omega \left(A^{(1)} -  \left< A^{(1)} \right> \right) \, dr'
\end{equation}
and 
\begin{equation}
Re(G^{(1)}_{y} \, \ell^{(0)*}_x) = - \left|\ell_x^{(0)} \right|^2   \int_{r_{\rm in}}^r \Sigma r'^3 \Omega \left(B^{(1)} -  \left< B^{(1)} \right> \right)\, dr',
\end{equation}
where $<>$ denotes the angular momentum weighted average, as was applied in Equation (\ref{javg}).
Using the fact the $ B^{(1)}(r)/ A^{(1)}(r)$ is independent of $r$, we obtain
that
\begin{equation}
\frac{d \chi^{(1)}}{dr} =0 
\label{chiconst}
\end{equation}
and therefore $\chi^{(1)}$ is constant in $r$.

\subsection{Evaluation  of  Tilt Component at Disc Inner Edge}

We consider $r$ times Equation (\ref{l1}) in second order. We integrate that equation over the entire disc and apply the boundary condition Equation (\ref{GBC}) to obtain
\begin{equation}
i \omega^{(2)} \ell^{(0)}_x  +  i \omega^{(1)} \left<  \tilde{\ell}^{(1)}_x  \right> =  \left<   A^{(1)} \right>  \ell^{(1)}_y(r_{\rm in})  + \left<   A^{(1)} \, \tilde{\ell}^{(1)}_y  \right>
\label{om2x} 
\end{equation}
and
\begin{equation}
i \omega^{(2)} \ell^{(0)}_y + i \omega^{(1)} \ell^{(1)}_y(r_{\rm in})  +  i \omega^{(1)} \left<  \tilde{\ell}^{(1)}_y  \right> = \left<   B^{(1)} \, \tilde{\ell}^{(1)}_x  \right> ,
\label{om2y}
\end{equation}
where
\begin{equation}
 \bm{\tilde{\ell}}^{(1)} (r)= \int^r_{r_{\rm in}} \frac{ 4 \alpha \Omega \bm{G}^{(1)}}{{\cal I} r'^3 \Omega^3} \, dr'.
 \label{eltil}
\end{equation}

We eliminate $\omega^{(2)}$ and solve for  $\ell^{(1)}_y(r_{\rm in})$ to obtain
\begin{equation}
\ell^{(1)}_y(r_{\rm in}) = \frac{ \left(\left<   B^{(1)} \, \tilde{\ell}^{(1)}_x   \right>   \ell^{(0)}_x 
- \left<   A^{(1)} \, \tilde{\ell}^{(1)}_y  \right>  \ell^{(0)}_y  \right)
+  \left(\left<   B^{(1)} \right> \, \left< \tilde{\ell}^{(1)}_x  \right>  \ell^{(0)}_x -\left<   A^{(1)} \right> \, \left< \tilde{\ell}^{(1)}_y  \right> \ell^{(0)}_y  \right)  }{2 \left<   A^{(1)} \right> \ell^{(0)}_y  }.
\label{ly1rin}
\end{equation}
By applying  Equations (\ref{G0}) and (\ref{eltil}) we have that 
\begin{equation}
\left<   B^{(1)} \, \tilde{\ell}^{(1)}_x  \right>  \ell^{(0)}_x =  
  \ell^{(0)}_x  \ell^{(0)}_y \int_{r_{\rm in}}^{r_{\rm out}} dr'''  \int_{r_{\rm in}}^{r'''} dr''
 \int_{r_{\rm in}}^{r''} dr'  f(r', r'', r''') 
  B^{(1)}(r''')   \left( A^{(1)}(r')  - \left<  A^{(1)} \right> \right) 
 \label{ell1p1}
\end{equation}
and 
\begin{equation}
\left<   A^{(1)} \, \tilde{\ell}^{(1)}_y  \right>  \ell^{(0)}_y =  
 \ell^{(0)}_x  \ell^{(0)}_y  \int_{r_{\rm in}}^{r_{\rm out}} dr'''  \int_{r_{\rm in}}^{r'''} dr''
 \int_{r_{\rm in}}^{r''} dr' f(r', r'', r''')  A^{(1)}(r''')  \left( B^{(1)}(r')  - \left<  B^{(1)} \right> \right)
  \label{ell1p2}
\end{equation}
where $f$ is a function that is dependent of $A^{(1)}$ and $B^{(1)}$.

Using the fact that $B^{(1)}(r)/ A^{(1)}(r)$ is independent of $r$ together with Equations (\ref{ell1p1}) and  (\ref{ell1p2}), we then have that
the first term in parenthesis on the RHS of Equation (\ref{ly1rin}) is zero.
It similarly follows that the second term in parenthesis is also zero.
Consequently,
\begin{equation}
\ell^{(1)}_y(r_{\rm in}) =0.
\label{ly1in}
\end{equation}

\subsection{Evaluation  of the Product Term at All Radii}

 Using Equations (\ref{lxin}) and (\ref{ly1in}), we then have that
\begin{equation}
\chi^{(1)}(r_{\rm in}) =  Re \left(\ell^{(0)}_x(r_{\rm in}) \ell^{(1)*}_y(r_{\rm in}) \right) =0.
\end{equation}
From Equation (\ref{chiconst}), we then have that
\begin{equation}
\chi^{(1)}(r) = 0 
\label{chi10}
\end{equation} 
for all $r$ in the disc.
It then follows that $Re(\ell_x(r)\, \ell_y^*(r))=0$ in first order and therefore the RHS of Equation (\ref{Tl2}) is zero.

\bsp	
\label{lastpage}
\end{document}